\documentclass[onecolumn,amsmath,amssymb]{qm2008_abs}
\usepackage{graphicx}
\usepackage{dcolumn}
\usepackage{bm}
\begin{document}

\title{{\Large Trace Anomaly and Dimension Two Gluon Condensate Above the Phase Transition\footnote{Presented by E.~Meg\'{\i}as at the 20th International Conference on Ultra-Relativistic Nucleus-Nucleus Collisions: Quark Matter 2008 (QM2008), Jaipur, India, 4-10 Feb 2008.} }}


\bigskip
\bigskip
\author{\large E.~Meg\'{\i}as}
\email{emegias@quark.phy.bnl.gov}
\affiliation{Physics Department, Brookhaven National Laboratory, 
Upton, New York 11973, USA}

\author{\large E.~Ruiz Arriola}
\email{earriola@ugr.es}

\author{\large L.L.~Salcedo}

\affiliation{Departamento de F\'{\i}sica At\'omica, Molecular y Nuclear, 
Universidad de Granada, 
E-18071 Granada, Spain}
\bigskip
\bigskip

\begin{abstract}
\leftskip1.0cm \rightskip1.0cm The dimension two gluon condensate has
been used previously within a simple phenomenological model to
describe power corrections from available lattice data for the
renormalized Polyakov loop and the heavy quark-antiquark free energy
in the deconfined phase of QCD~\cite{Megias:2005ve,Megias:2007pq}. The
QCD trace anomaly of gluodynamics also shows unequivocal inverse
temperature power corrections which may be encoded as dimension two
gluon condensate. We analyze lattice data of the trace anomaly and
compare with other determinations of the condensate from previous
references, yielding roughly similar numerical values.
\end{abstract}

\maketitle

{\bf Introduction.} For zero and for infinite quark masses (gluodynamics)
QCD is invariant under scale and conformal transformations at the
classical level. This classical invariance is broken, however, by
quantum corrections due to the necessary regularization of ultraviolet
divergences which introduces a mass scale, $\Lambda_{\rm QCD}$; the
divergence of the dilatation current equals the trace of the improved
energy-momentum tensor $\Theta^\mu_\mu$~\cite{Callan:1970ze} yielding
the so-called {\it ``trace anomaly''}~\cite{Collins:1976yq}. At finite
temperature, the energy density $\epsilon$ and the pressure $p$
enter as~\cite{Landsman:1986uw,Ellis:1998kj,Drummond:1999si,Agasian:2001bj},
\begin{equation}
\epsilon - 3p = \frac{\beta(g)}{2g} \langle (G^a_{\mu\nu})^2\rangle \equiv \langle \Theta^\mu_\mu\rangle  \,,
\label{eq:tr_an}
\end{equation}
where $G_{\mu\nu} = \partial_\mu A_\nu - \partial_\nu A_\mu + ig
[A_\mu,A_\nu]$ is the field strength tensor and $\beta(g) = \mu
\partial g / \partial \mu $ is the beta function. Far from the
conformal limit, where $\epsilon = 3p$, $\Delta =(\epsilon - 3 p
)/T^4$ is a dimensionless quantity providing a measure of the
interaction, so it is commonly known as {\it ``interaction
measure''}. A good knowledge of $\Delta$ is crucial to understand the
deconfinement process, where the non perturbative (NP) nature of low
energy QCD seems to play a prominent role. In this contribution we
analyze the highly NP behaviour of the trace anomaly just above the
phase transition and describe it in a way that is consistent with
other thermal observables (see~\cite{Megias:2008ip} for further
details).

\medskip

\begin{figure}[tbp]
\begin{center}
\begin{minipage}[t]{7.3cm}
\includegraphics[width=7.3cm]{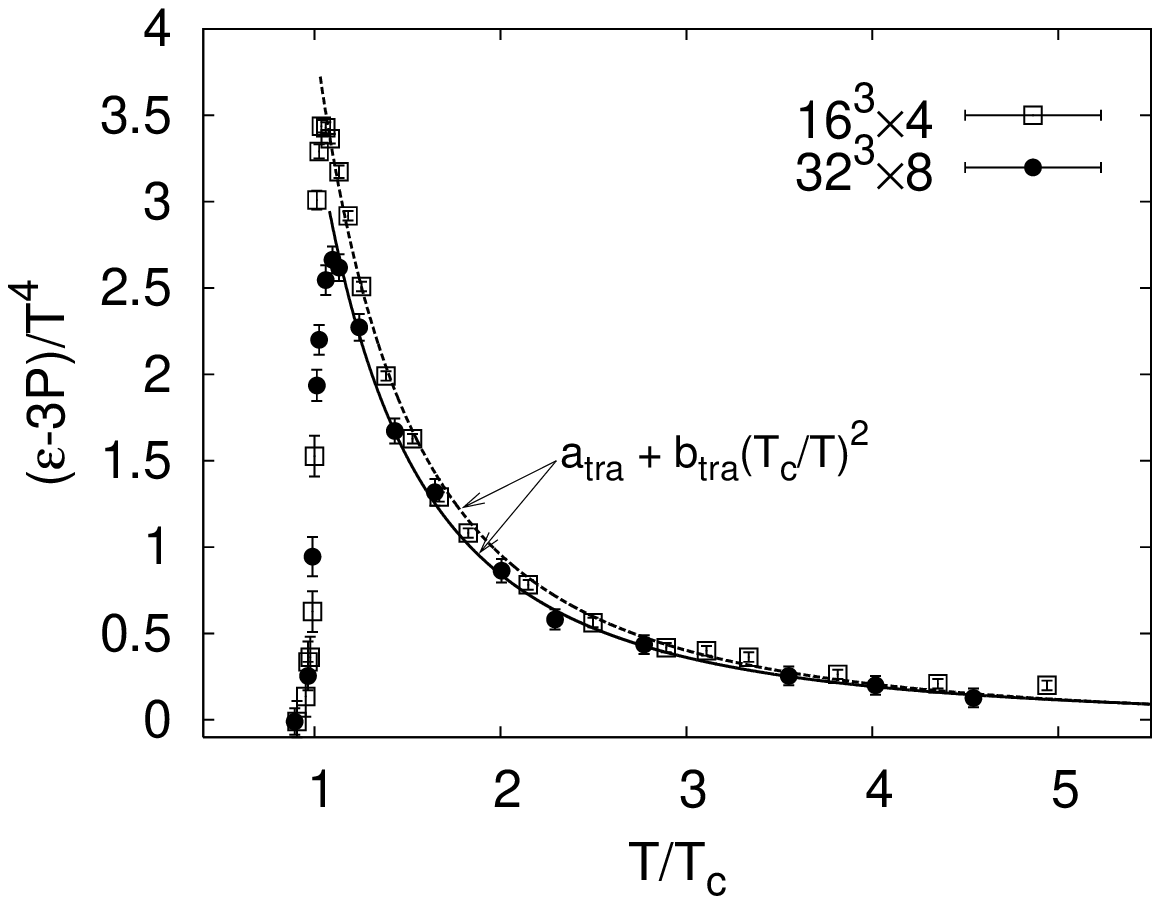}
\end{minipage}
\begin{minipage}[t]{7.3cm}
\includegraphics[width=7.3cm]{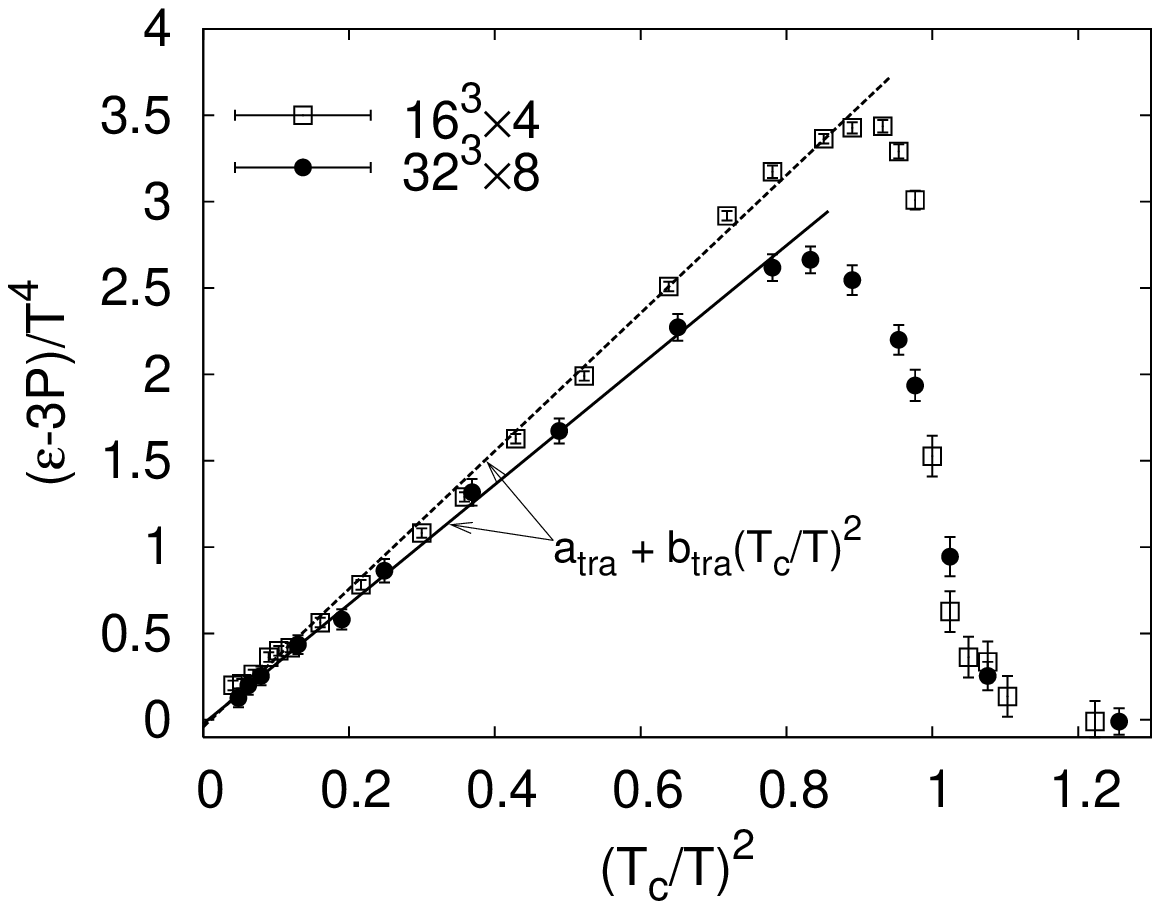}
\end{minipage}
\end{center}
\caption{The trace anomaly density $ (\epsilon -3 P)/T^4 $ as a function of $T$ (left) and $1/T^2$ (right) (in units of $T_c$). Lattice data are from~\cite{Boyd:1996bx} for $N_\sigma^3 \times N_\tau = 16^3 \times 4$ and
$32^3 \times 8$. The fits use Eq.~(\ref{eq:e3pfit}) with $a_{\rm tra}$ and 
$b_{\rm tra}$ adjustable constants.}
\label{fig:e3p}
\end{figure}

{\bf Thermal power corrections in gluodynamics.} The interaction
measure was computed one decade ago on the lattice by the Bielefeld
group for gluodynamics~\cite{Boyd:1996bx}. Fig.~\ref{fig:e3p} shows
the lattice data for $\Delta =(\epsilon - 3p)/T^4$ as a function of
$T/T_c$.  $\Delta$ is very small below $T_c$, because the lightest glueball
is much heavier than $T_c \approx 270\, {\rm MeV}$.  It increases
suddenly near and above $T_c$ by latent heat of deconfinement, and
raises a maximum at $T \approx 1.1\, T_c$. Then it has a gradual
decrease reaching zero in the high temperature limit. The high value
of $\Delta$ for $ T_c \lesssim T \lesssim (2.5-3) T_c$ corresponds to
a strongly interacting Quark-Gluon Plasma picture.

From our previous experience~\cite{Megias:2005ve,Megias:2007pq} and
following a remark by Pisarski~\cite{Pisarski:2006yk}, in
Fig.~\ref{fig:e3p} we plot $(\epsilon - 3p)/T^4$ as a function of
$1/T^2$ (in units of $T_c$) exhibiting an unmistakable straight line
behaviour in the region slightly above the critical temperature, of the form 
\begin{equation}
\Delta=(\epsilon - 3 P)/T^4 = 
a_{\rm tra} + b_{\rm tra} \left(T_c /T \right)^2  \,, 
\label{eq:e3pfit} 
\end{equation}  
and  corresponding to a ``power correction'' in temperature. 
A fit of the lattice data ($N_\sigma^3 \times N_\tau = 32^3 \times 8$)
for $T/T_c > 1.13$ yields $a_{\rm tra} = -0.02(4)\,,\; b_{\rm tra} =
3.46(13)\,, \; \chi^2/ {\rm DOF} = 0.35 \,$. Power corrections also
appear in $\epsilon$ and $P$, just by applying the standard
thermodynamic relations.

\begin{figure}[tbp]
\begin{center}
\begin{minipage}[t]{7.3cm}
\includegraphics[width=7.3cm]{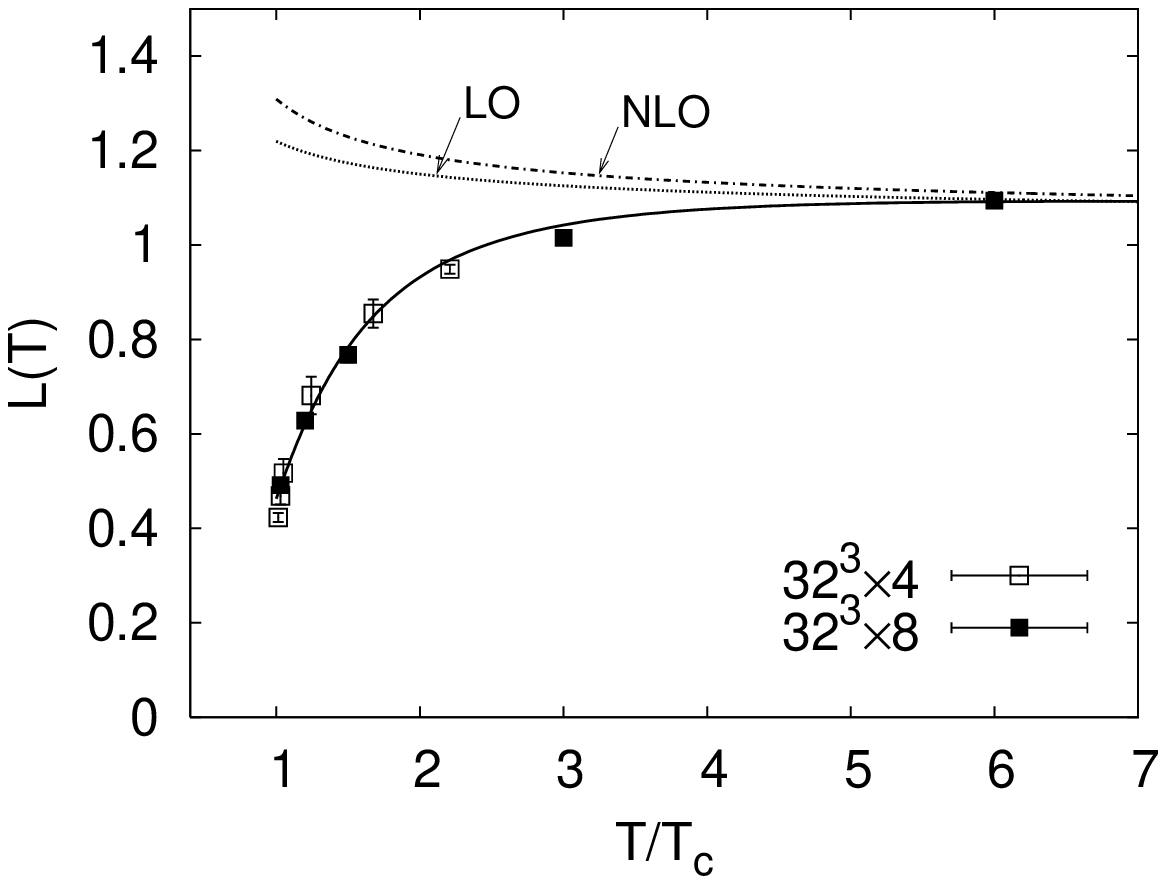}
\end{minipage}
\begin{minipage}[t]{7.3cm}
\hspace{0.0cm}\includegraphics[width=7.3cm]{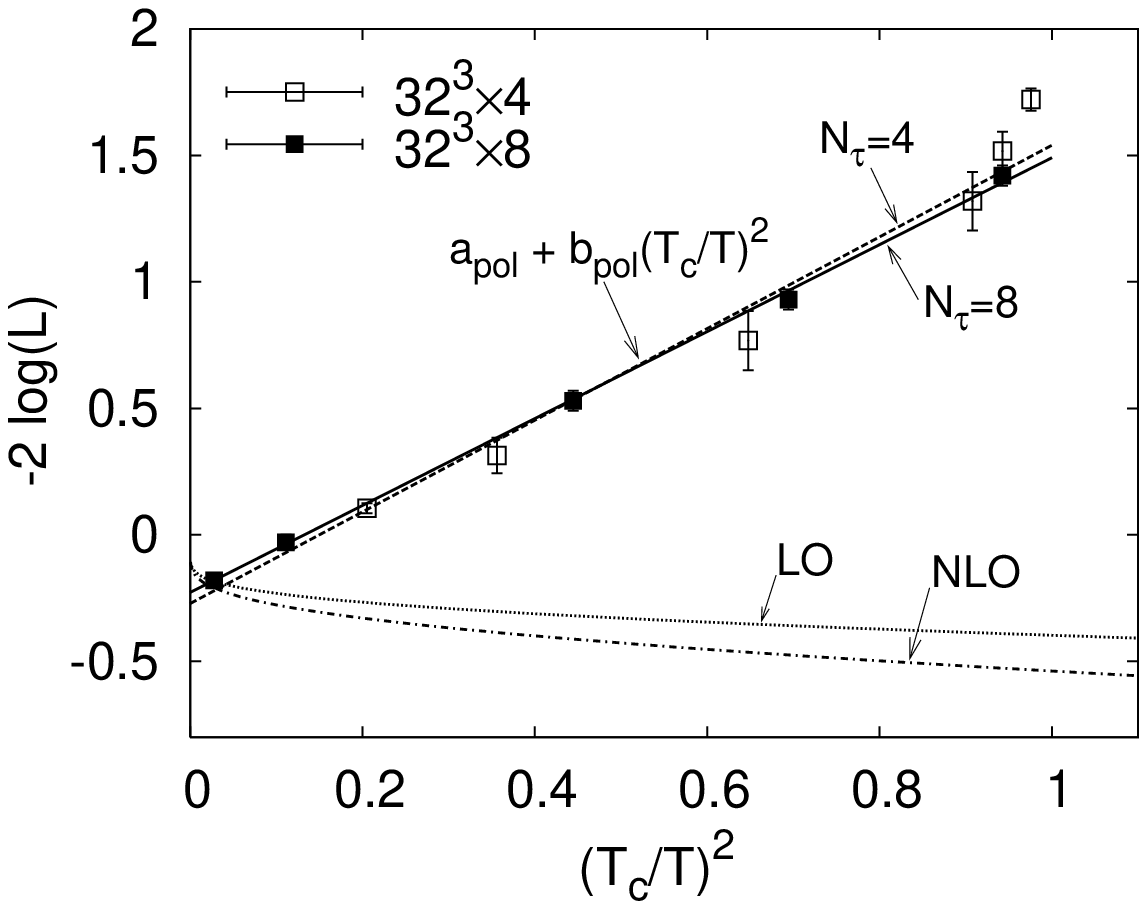}
\end{minipage}
\end{center}
\caption{The temperature dependence of the renormalized Polyakov loop
in units of the critical temperature. Lattice data are
from~\cite{Kaczmarek:2002mc}, for $N_\sigma^3 \times N_\tau = 32^3
\times 4$ and $32^3 \times 8$. We plot the perturbative result at LO
and NLO, and the fit using $a_{\rm pol} + b_{\rm pol}\left( T_c/T \right)^2 $.}
\label{fig:PL}
\end{figure}

This behaviour clearly contradicts perturbation theory (PT) which
contains no powers but only logarithms in the temperature, a feature
shared by hard thermal loops and other resummation techniques (see
e.g.~\cite{Andersen:1999sf,Andersen:2004fp}), explaining why they have
failed to describe lattice data of the free energy below $3 T_c$. 
Fig.~\ref{fig:PL} shows the lattice data of Polyakov loop from
Ref.~\cite{Kaczmarek:2002mc}, suggesting
again~\cite{Megias:2005ve,Megias:2007pq} a linear fit of the form $-2
\log L = a_{\rm pol} + b_{\rm pol} \left( T_c/T \right)^2 \,$. In what
follows, we show a phenomenological model that describes consistently
all these power corrections in an unified way.

\medskip

{\bf Dimension two gluon condensate.} The gluon condensate $\langle
G^2 \rangle \equiv g^2 \langle (G^a_{\mu\nu})^2\rangle $ describes the
anomalous (and not spontaneous) breaking of scale invariance, and
hence is not an order parameter of the phase transition. Actually, the
order parameter is the vacuum expectation value of the Polyakov loop
which signals the breaking of the ${\mathbb Z}(N_c)$ discrete symmetry
of gluodynamics as well as the deconfinement transition. A
dimension two gluon condensate naturally appears from a computation of
the Polyakov loop in which a Gaussian distribution of eigenvalues is
considered. In the static gauge, $\partial_0 A_0 ({\mathbf x},x_0)=0$,
this Gaussian-like, large $N_c$ motivated, approximation
gives~\cite{Megias:2005ve}
\begin{equation}
L(T) = \left\langle \frac{1}{N_c} \, {\rm tr}_c \, e^{i g A_0({\mathbf x})/T } \right\rangle = 
\exp \left[  -\frac{g^2 \langle A_{0,a}^2 \rangle }{4N_c T^2} \right] + {\cal O}(g^6) \,,
\label{eq:defPL}
\end{equation}
valid up to ${\cal O}(g^5)$ in PT. $A_0$ is the gluon field in the
(Euclidean) time direction. From here it is immediate to relate the
Polyakov loop to the gluon propagator in the dimensionally reduced
theory
\begin{equation}
\delta_{ab} T  D_{00}({\mathbf k}) = \int d^3 {\mathbf x}
\langle A_{0,a}({\mathbf x}) A_{0,b}({\mathbf y})\rangle  e^{-i {\mathbf k} \cdot ({\mathbf
x}-{\mathbf y})}  \,.
\label{eq:dim2gc}
\end{equation}
The dimension two gluon condensate $g^2\langle A_{0,a}^2 \rangle$ is
obtained from Eq.~(\ref{eq:dim2gc}) in the limit ${\mathbf x}
\rightarrow {\mathbf y}$. The perturbative propagator $D_{00}^{\rm
P}({\mathbf k}) = 1/({\mathbf k}^2 + m_D^2) + {\cal O}(g^2)$, being
$m_D \sim T$ the Debye mass, leads to the known perturbative result of
Gava and Jengo~\cite{Gava:1981qd}, which fails to reproduce lattice
data below $6 T_c$. A NP model is proposed in
Ref.~\cite{Megias:2005ve} to describe the lattice data of the Polyakov
loop, and it consists in a new piece in the gluon propagator driven by
a positive mass dimension parameter:
\begin{equation}
D_{00}({\mathbf k}) = D_{00}^{\rm P}({\mathbf k}) + D_{00}^{\rm NP}({\mathbf k}) \,, \qquad D_{00}^{\rm NP}({\mathbf k}) =m_G^2 /({\mathbf k}^2 + m_D^2)^2 \,.
\label{eq:NPmodel}
\end{equation}
This ansatz parallels a zero temperature one~\cite{Chetyrkin:1998yr},
where the dimension two condensate provides the short-distance NP
physics of QCD and  at zero temperature this contribution yields the well
known NP linear term in the $\overline{q}q$ potential. 
A justification of Eq.~(\ref{eq:NPmodel}) based on Schwinger-Dyson
methods has been given~\cite{Gogokhia:2005gk}. The new propagator
generates a NP contribution to the condensate, $\langle
A_{0,a}^2\rangle = \langle A_{0,a}^2\rangle^{\rm P} + \langle
A_{0,a}^2\rangle^{\rm NP}$, which is related to $m_G^2$ through
$\langle A_{0,a}^2\rangle^{\rm NP} = (N_c^2-1) m_G^2/(8\pi {\hat
m}_D)$, where ${\hat m}_D \equiv m_D/T$, so that it leads to the
thermal power behaviour that we observe in Fig.~\ref{fig:PL}. The
Gaussian approximation has also been used in Ref.~\cite{Megias:2007pq}
to compute the singlet free energy of a heavy ${\overline q}q$
pair~\cite{Kaczmarek:2002mc,Kaczmarek:2004gv}, through the correlation
function of Polyakov loops.

\medskip

{ \bf Non perturbative contribution to the Trace Anomaly.} The model
of Eq.~(\ref{eq:NPmodel}) can be easily used to compute the trace
anomaly Eq.~(\ref{eq:tr_an}) in gluodynamics. Assuming the leading NP 
contribution to
be encoded in the $A_{0,a}$ field and taking $A_{i,a}=0$ yields
\begin{equation}
\langle G^a_{\mu\nu} G^a_{\mu\nu}\rangle^{\rm NP} = 2 \langle \partial_i A_{0,a} \partial_i A_{0,a}\rangle^{\rm NP} = -6 m_D^2 \langle A_{0,a} A_{0,a}\rangle^{\rm NP}\,. 
\end{equation}  
The r.h.s. is obtained from Eq.~(\ref{eq:dim2gc}) by expanding in the
limit ${\mathbf x} \rightarrow {\mathbf y}$ and looking at the
quadratic term in $r = |{\mathbf x}-{\mathbf y}|$. Note that the NP
model is formulated in the dimensionally reduced theory, so the gluon
fields are static. This formula produces the thermal power behaviour
of Eq.~(\ref{eq:e3pfit}) with
\begin{equation}
b_{\rm tra} T_c^2 = -3 \hat{m}_D^2  \langle A_{0,a}^2\rangle^{\rm NP} \beta(g) /g \,. 
\label{eq:btra}
\end{equation}
 If we consider the perturbative value of the beta function $\beta(g)
 \sim g^3 + {\cal O}(g^5) $, the r.h.s. of Eq.~(\ref{eq:btra}) shows a
 factor $g^2$ in addition to the dimension two gluon condensate $g^2
 \langle A_{0,a}^2\rangle^{\rm NP}$. So the fit of the trace anomaly
 data is sensitive to the value of the smooth $T$-dependent $g$, without
 jeopardizing the power correction. For the Polyakov loop the
 sensitivity in $g$ is only through the perturbative terms, which are
 much smaller than the NP ones. When we consider the perturbative
 value $g_P$ up to 2-loops, we get from the fit of the trace anomaly
 $g^2 \langle A_{0,a}^2\rangle^{\rm NP} = (2.63 \pm 0.05 \, T_c )^2$,
 which is a factor $1.5$ smaller than what is obtained from other
 observables. This disagreement could be partly explained on the basis
 of certain ambiguity of $g$ in the NP regime. A better fit of the
 Polyakov loop and heavy quark free energy lattice data in the regime
 $T_c < T < 4 T_c$ is obtained for a slightly smaller $g$ than $g_P$,
 i.e. $g = 1.26 - 1.46$~\cite{Megias:2007pq}. Taking this value we get
 from Eq.~(\ref{eq:btra}) $g^2 \langle A_{0,a}^2 \rangle^{\rm NP} =
 (2.86 \pm 0.24 \, T_c)^2$, in better agreement with other
 determinations, see Table~\ref{tab:1}. Nonetheless, an alternative
 method to compute the trace anomaly based on the direct
 computation of $\epsilon - 3p$ from the partition function of
 gluodynamics does reproduce the power correction,
 however with different coefficients~\cite{Megias:2008ip}.

\begin{table}[htb]
\begin{center}
\begin{tabular}{|c|c|}
\hline
{\bf Observable} &  $\hspace{0.3cm} {\bf g^2 \langle A_{0,a}^2 \rangle^{\rm NP} }$ \hspace{0.3cm}  \\
\hline
\hspace{0.3cm} Polyakov loop~\cite{Megias:2005ve} \hspace{0.3cm} &  \hspace{0.3cm} $(3.22 \pm 0.07 \, T_c)^2   $ \hspace{0.3cm} \\
\hspace{0.3cm} Heavy $\overline{q}q$ free energy~\cite{Megias:2007pq} \hspace{0.3cm}  &  \hspace{0.3cm} $(3.33 \pm 0.19 \, T_c)^2$  \hspace{0.3cm} \\ 
\hspace{0.3cm} Trace Anomaly  \hspace{0.3cm} &   \hspace{0.3cm} $(2.86 \pm 0.24 \, T_c)^2$ \hspace{0.3cm} \\ 
\hline
\end{tabular}
\end{center}
\caption{Values of the dimension two gluon condensate from a fit of
several observables in the deconfined phase of gluodynamics: Polyakov
loop, singlet free energy of heavy quark-antiquark and trace
anomaly. Values are in units of $T_c$. We show the fit for lattice
data with $N_\tau = 8$. Error in last line takes into account an
indeterminate value of the coupling constant $g = 1.26 - 1.46$, being
the highest value the perturbative $g_P$ up to 2-loops at $T = 2
T_c$. The critical temperature in gluodynamics is $T_c = 270 \pm 2 \,$MeV~\cite{Kaczmarek:2002mc}.}
\label{tab:1}
\end{table}

\medskip

{ \bf Summary and conclusions.} The trace anomaly in gluodynamics
shows, near and above the critical temperature, a clear pattern of
power corrections which cannot be matched to perturbation theory or
resummations thereof. It can instead be explained in terms of a
dimension two gluon condensate whose numerical value agrees with other
determinations based on other thermal observables and it is also
remarkably close to existing studies at $T=0$ (see
e.g. Refs.~\cite{Boucaud:2001st,RuizArriola:2004en}). 

\medskip
{\bf Acknowledgments.} We thank R.D.~Pisarski for correspondence.
E.~Meg\'{\i}as is supported from the joint sponsorship by the
Fulbright Program of the U.S. Department of State and Spanish Ministry
of Education and Science.  Work supported by Spanish DGI and FEDER
funds with grant FIS2005-00810, Junta de Andaluc\'{\i}a grant
FQM-225-05, EU Integrated Infrastructure Initiative Hadron Physics
Project contract RII3-CT-2004-506078, and U.S. Department of Energy
contract DE-AC02-98CH10886.

\end{document}